\documentclass[12pt,showpacs,aps,preprintnumbers,prl,amsmath,amssymb]{revtex4}
\usepackage{graphicx}
\usepackage{epsfig,color}

\begin{document}

\title{Pressure-induced spin-density-wave transition in superconducting Mo$_3$Sb$_7$}
\author{V. H. Tran$^1$, R. T. Khan$^2$, P. Wi{\'s}niewski$^1$, E. Bauer$^2$}
\affiliation{$^1$ Institute of Low Temperature and Structure Research, Polish Academy of Sciences, P. O. Box 1410, 50-950 Wroc\l aw, Poland\\
$^2$  Institute of Solid State Physics, Vienna University of Technology, A-1040 Wien, Austria
}

\begin{abstract}
We report a novel pressure-induced spin-density-wave transition in the superconductor Mo$_3$Sb$_7$ figured our by measurements of the electrical resistivity and magnetization under hydrostatic pressure. The critical temperature of superconducting Mo$_3$Sb$_7$ is found to increase with increasing pressure, from  2.15 K at 0.2 kbar up to 2.37 K at 22 kbar. Above 4.5 kbar, superconductivity exists in parallel with a pressure-induced spin-density wave state, revealed by a sharp jump in the electrical resistivity and a maximum in the magnetization at the phase transition temperature $T_{SDW}$. The application of pressure shifts $T_{SDW}$ to lower temperatures, from 6.6 K at 4.5 kbar down to 6.15 K at 22 kbar. A strong magnetic field dependence of $T_{SDW}$ and a maximum seen in the magnetization indicate an antiferromagnetic character of $T_{SDW}$. The pressure dependence of $T_c$ and $T_{SDW}$ suggests a competition of the SDW and the superconducting states in this system.
\end{abstract}

\maketitle
\par Mo$_3$Sb$_7$ crystallizes in the cubic Ir$_3$Ge$_7$-type structure with space group Im-3m and \emph{a} = 0.9591 nm. This compound has been suggested to enter into an unusual superconducting state below $T_c$ = 2.2 K with two BCS-like gaps.\cite{Tran_AM,TranETS} Moreover, specific heat and muon spin rotation experiments,\cite{Tran08b} indicated that the effective mass of the charge carries is enhanced ($\sim$ 16-18 $m_0$), if compared to that of conventional BCS superconductors with a weak electron-phonon coupling. At present, Mo$_3$Sb$_7$ is concluded being nonmagnetic due to a spin dimerization below $T^*$ = 50 K, which is reflected by anomalies in the heat capacity, magnetic susceptibility\cite{Tran08}, muon spin rotation,\cite{Tran09,Koyama,Tabata} and magnetic excitations in inelastic neutron scattering.\cite{Tran09} The crystal structure of this compound is certainly favorable for spin dimerization, because of a distinct difference between the intradimer distance ($\sim$ 0.3 nm) and interdimer distance ($\sim$ 0.46 nm). In general, the strength of magnetic interactions in a given compound can be modified by applying external hydrostatic pressure. Furthermore, taking into account the fact that the effective mass of superconducting carriers in Mo$_3$Sb$_7$ is sizable, approaching that of heavy-fermion superconductors, one would expect a similar response of its electronic state to applied pressure. In heavy-fermion superconductors, magnetic instabilities or magnetic fluctuations are responsible for enhancing the effective mass of the superconducting quasiparticles and the pairing mechanism in unconventional superconductors is most probably associated with magnetism.\cite{Mathur,Sato,Moriya,Coleman} In view of these arguments, a considerable pressure response on the magnetic state is expected in  Mo$_3$Sb$_7$. Therefore, the electrical resistivity and magnetization of Mo$_3$Sb$_7$ was studied under pressure. Results of our study revealed that a pressure-induced spin-density-wave exists in parallel with superconductivity. In the following, we present pressure-dependent properties of Mo$_3$Sb$_7$ and discuss these observation in the context of a competition of magnetism and superconductivity.
\par Mo$_3$Sb$_7$ was prepared from Mo and Sb (purity 99.95$\%$ from Alfa Aesar) by solid-state reaction. Synthesis, purity characterization and determination of crystal parameters of the sample have been carried out in a similar technique as given in Ref. [\onlinecite{tran08a}]. Electrical resistivity under pressure up to 22 kbar was measured in the temperature range 0.3-80 K by means of a standard four-probe technique, using a $^3$He cryostat with a Cryogenics Ltd. 12 T magnet. Hydrostatic pressure was generated by a piston-cylinder cell using Daphne oil as the pressure-transmitting medium. The resistivity data were taken on cooling at a rate below 0.1 K/min. The magnetization was measured at several pressures up to 6 kbar in a nonmagnetic pressure cell with a SQUID magnetometer (Quantum Design MPMS-5). High purity Pb was used as pressure indicator. The width of the superconducting transition of  Pb did not exceed 10 mK, corresponding to an uncertainty of measured pressure $\pm$ 0.2 kbar.
\par
\begin{figure}[h]
\includegraphics[scale=0.75]{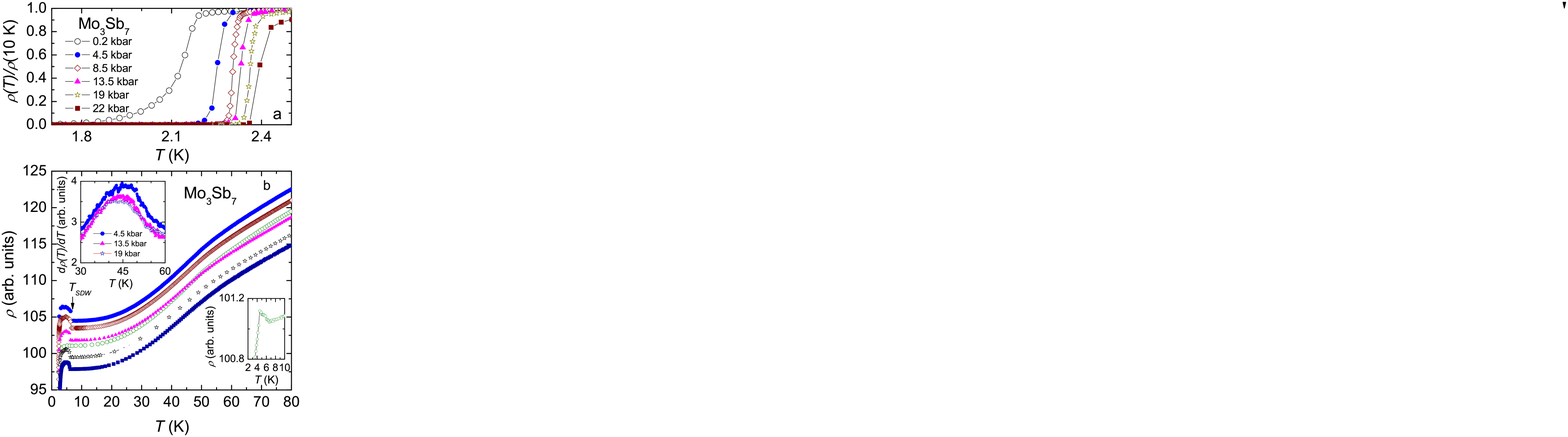}
\caption{ (color online) Electrical resistivity of Mo$_3$Sb$_7$ for different pressures. a) The low-temperature data are normalized at T = 10 K. The superconducting transition is very sharp only under pressure. b) Electrical resistivity of Mo$_3$Sb$_7$ at temperatures up to 80 K. The upper inset shows temperature derivative of the resistivity vs. temperature. The bottom inset shows the resistivity at 0.2 kbar for temperature range 4 -10 K.}
\label{fig:Fig_1}
\end{figure}

 The  normalized low temperature electrical resistivity of Mo$_3$Sb$_7$ for various pressures up to 22 kbar is presented in Fig. \ref{fig:Fig_1} a. The resistivity at 0.2 kbar is characterized by a superconducting transition at 2.15 K, in agreement with the previous report for ambient pressure.\cite{bukowski} With growing pressure, the critical temperature increases at an initial rate $dT_c/dP \sim $ 0.02 K/kbar, and at maximum applied pressure of 22 kbar $T_c$ reaches a value of 2.37 K. Simultaneously, the width of the critical transition $\Delta T_c$ decreases, from 0.2 K at 0.2 kbar to 0.1 K at 22 K, indicating that a more homogeneous superconducting state is realized under pressure. The observations of an increase of $T_c$ and of a gradually sharpening of the phase transition with increasing pressure (see Fig. \ref{fig:Fig_1} a) imply that superconductivity is favored by pressure. This behavior cannot be explained in terms of the BCS theory for conventional superconductors like Al, Pb or Nb$_3$Sn. For  BCS-superconductors, the critical temperature is given by: $k_BT_c \sim  \omega_D exp(\frac{-1}{N(E_F)V_{e-ph}})$,\cite{bcs} thus $T_c$ is governed by the phonon frequency $\omega_D$, the electron-phonon interaction $V_{e-ph}$ and the carrier density of states at the Fermi energy $N(E_F)$. Usually, $T_c$ is found to decrease with increasing pressure due to a lattice stiffening and a decrease of both the density of states and of $V_{e-ph}$ under pressure.\cite{aaa,bbb,ccc} Previously, pressure-enhanced superconductivity has been observed in high-$T_c$$^{\prime}s$ superconductors like YBa$_2$Cu$_4$O$_8$,\cite{ddd} HgBa$_2$Ca$_{m-1}$Cu$_m$O$_{2m+2+\delta}$\cite{Gao}, Bi$_2$Sr$_{1.5}$La$_{0.5}$CuO$_{6+\delta}$ or Bi$_2$Sr$_2$CaCu$_2$O$_{8+\delta}$,\cite{eee} and has been assigned to an increasing charge carrier density or to increasing interlayer coupling. A large enhancement of $T_c$ by pressure was found for clathrate Ba$_6$Ge$_{25}$ as well,\cite{Grosche} where under pressure the compound approaches the undistorted, low-disorder structure. This gives rise to a softening of phonon modes associated with one of the Ba atoms.  A possible explanation for Mo$_3$Sb$_7$ is based on the assumption that the electron pairing mechanism is mediated by antiferromagnetic interactions, as it will be discussed below.

\begin{figure}[h]
\includegraphics[scale=0.75]{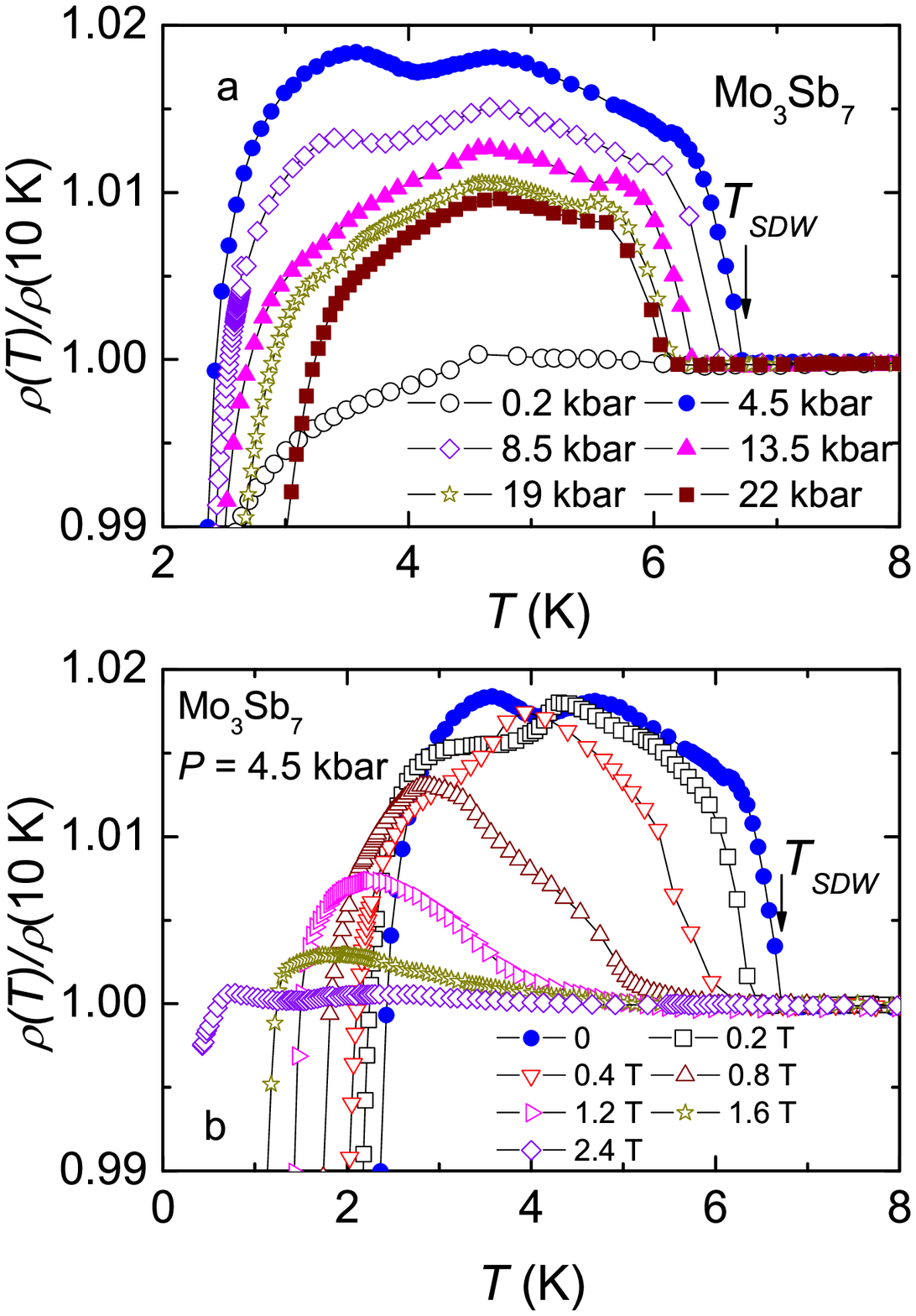}
\caption{ (color online) a) Electrical resistivity of Mo$_3$Sb$_7$ at zero field in the temperature range 2 - 8 K for different pressures.  b) Electrical resistivity of Mo$_3$Sb$_7$ at 4.5 kbar in the temperature range 0.3 - 8 K in several selected magnetic fields. The data are normalized at T = 10 K. The onset of the resistivity upturn is assumed to be the density-wave transition denoted as $T_{SDW}$.}
\label{fig:Fig_2}
\end{figure}

\par Fig. \ref{fig:Fig_1}b shows resistivity $\rho(T)$ data under pressure. Above about 7 K, Mo$_3$Sb$_7$ behaves metallic, without substantial changes upon a change of pressure. For temperatures around $T^*$, a shoulder in $\rho(T)$ becomes unveiled. We have analyzed the data around $T^*$ by taking the temperature derivative of the resistivity $d\rho(T)/dT$; results are plotted in the upper inset of Fig. \ref{fig:Fig_1} b. The maximum of $d\rho(T)/dT$, associated with the spin dimerization transition, decreases with increasing pressure. The most important finding of this work, however, is the observation of a pressure-induced phase transition as evidenced by a sharp anomaly of the 4.5 kbar  resistivity at 6.6 K, indicated by an arrow in Fig. \ref{fig:Fig_1} b and shown  in more detail in Fig. \ref{fig:Fig_2} a.  Nevertheless, a very tiny anomaly around 6.7 K can be recognized for the data collected at 0.2 kbar (see bottom inset of \ref{fig:Fig_1} b). It is reasonable to ascribe the upturn in the resistivity to the occurrence of a density wave phase, alike the spin-density wave in URu$_2$Si$_2$,\cite{Palstra} and BaFe$_{2-x}$Co$_x$As$_2$, \cite{SDW_SC} or the charge-density wave (CDW) in Lu$_5$Ir$_4$Si$_{10}$ \cite{shelton1} and Lu$_5$Rh$_4$Si$_{10}$. \cite{Yang} We attribute the resistivity upturn in Mo$_3$Sb$_7$ to the opening of an energy gap at some portions of the Fermi surface associated with a pressure-induced antiferromagnetic state. If this scenario is appropriate, the height of the resistivity jump may reflect the magnitude of the energy gap and obviously, the phase transition at $T_{SDW}$ becomes suppressed by higher applied pressures or magnetic fields. Indeed, with increasing pressure, $T_{SDW}$ shifts to lower temperatures and at the largest pressure applied (22 kbar) $T_{SDW} \sim $ 6.1 K. It is worth noting that the pressure data manifest complex $\rho(T)$-curves for $T_c < T < T_{SDW}$. The cause of several local maxima of $\rho(T)$ in this temperature range is unknown. Also, we examined the influence of a magnetic field on $T_{SDW}$. As an example, data at \emph{P} = 4.5 kbar and magnetic fields up to 2.4 T are shown in Fig. \ref{fig:Fig_2} b. Obviously, applied magnetic fields displace $T_{SDW}$ to lower temperatures. This behavior is different from that expected for a CDW-type transition, where the magnetic fields hardly influence the transition temperature.\cite{Jung} On the other hand, the strong sensitivity of $T_{SDW}$ in Mo$_3$Sb$_7$ to external magnetic fields is consistent with an antiferromagnetic SDW nature of the transition. The effect of magnetic fields on the transition is also manifested by a change of the resistivity slope. At zero field, a sharp jump appears in the resistivity at 4.5 kbar, suggesting a first-order phase transition. For higher magnetic field strengths, the transition becomes smoother, reminiscent of a second-order phase transition.
\par
\begin{figure}[h]
\includegraphics[scale=0.75]{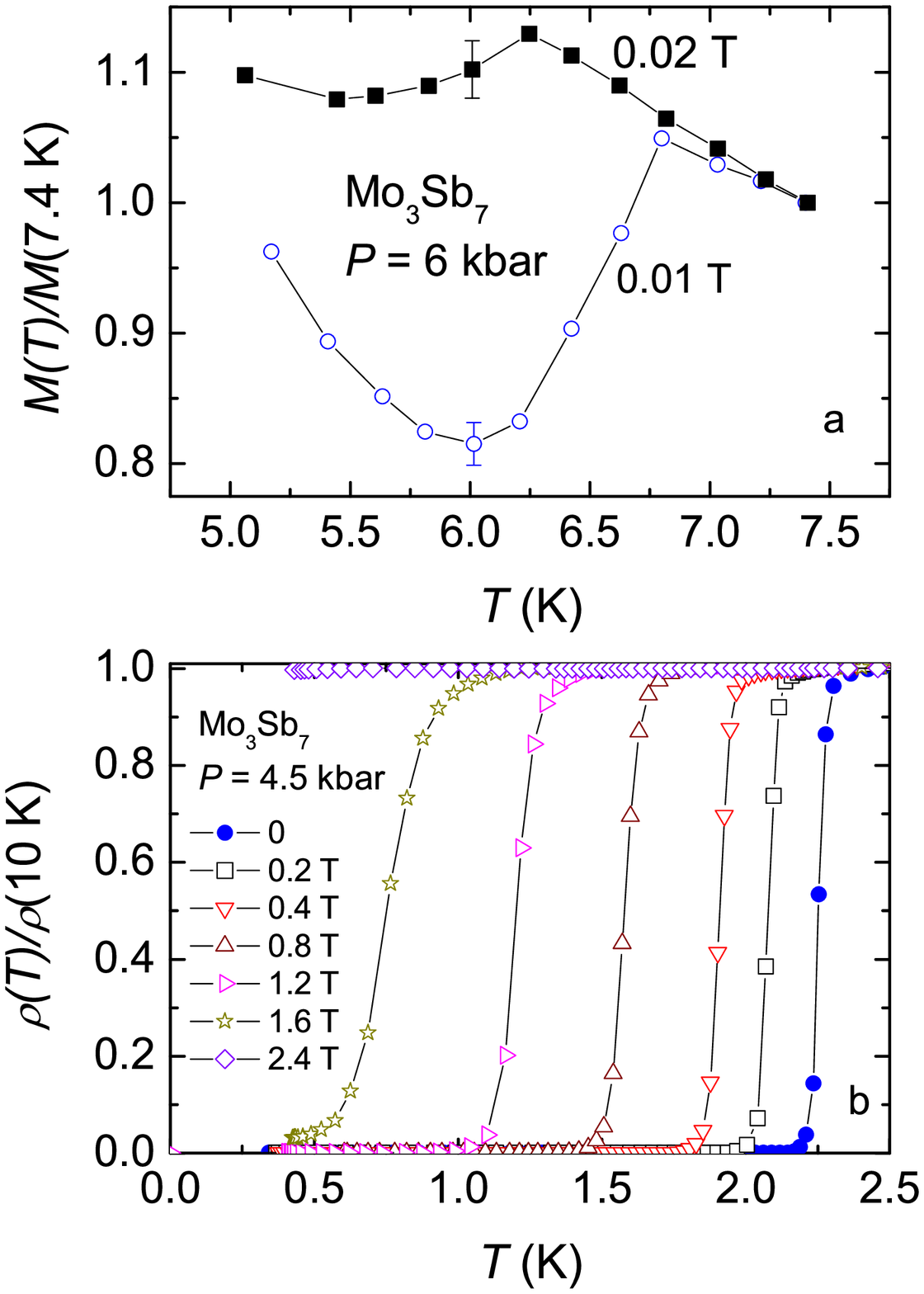}
\caption{\label{fig:Fig_3} (color online) a) dc-magnetization of Mo$_3$Sb$_7$ measured at 0.01 and 0.02 T in the temperature range 5 - 8 K. The data are normalized to the value at \emph{T} = 7.4 K. b) Electrical resistivity at 4.5 kbar in several selective magnetic fields.}
\end{figure}

 To corroborate the magnetic origin of the transition at $T_{SDW}$ the magnetization was measured in low magnetic fields up to 0.02 T and under pressure of 6 kbar (Fig. \ref{fig:Fig_3} a). While no anomalies occur at ambient pressure,\cite{Tran08} the magnetization under applied pressure and at 0.01 T exhibits a maximum near 6.8 K, referring to a pressure-induced antiferromagnetic phase transition. In a manner typical for antiferromagnets, the application of higher magnetic fields shifts the magnetization maximum to lower temperatures. According to the band calculations,\cite{Tran08} the Fermi surface is nested. Such a property may trigger a SDW ordering.
\par
\begin{figure}[h]
\includegraphics[scale=0.75]{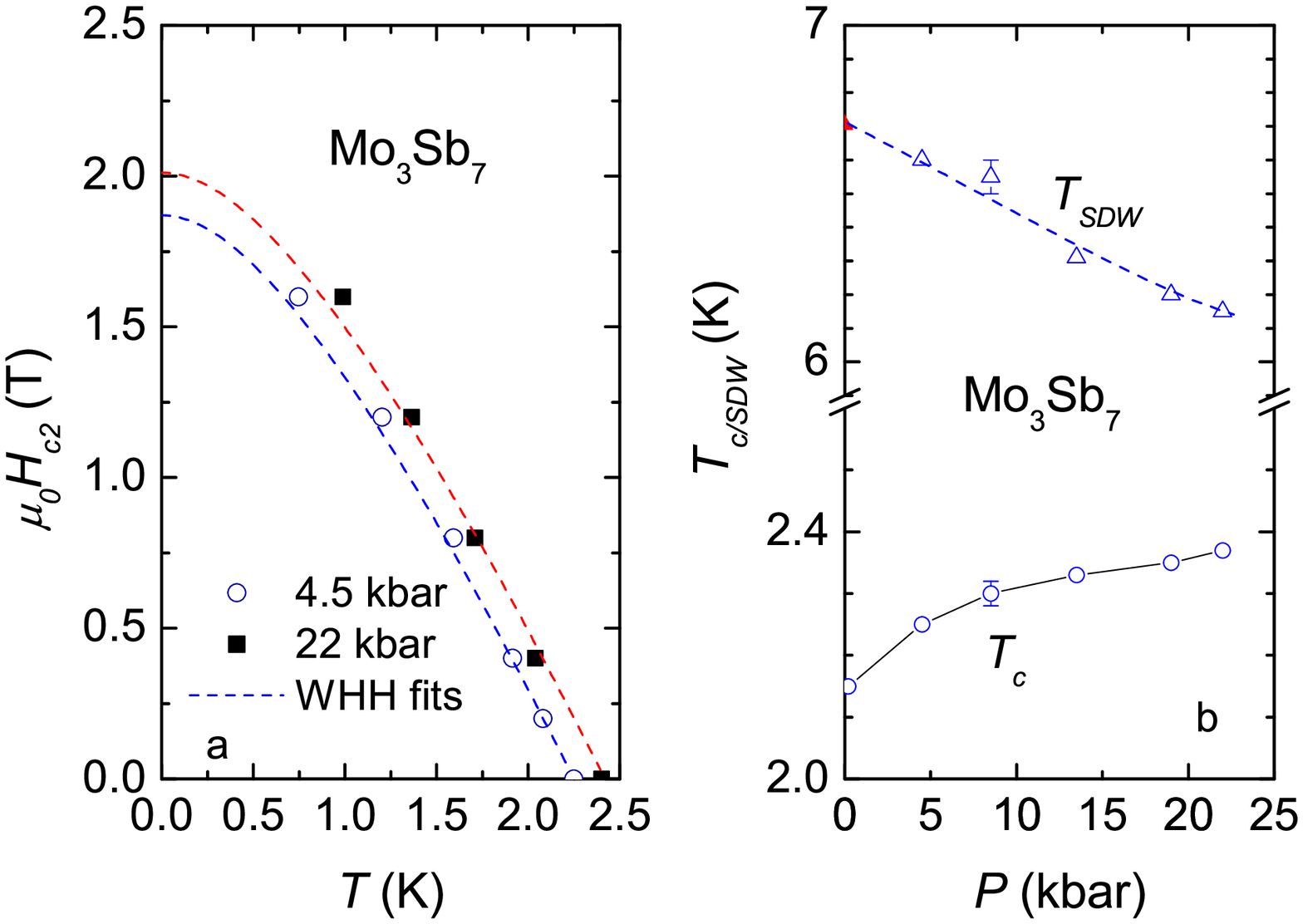}
\caption{\label{fig:Fig_4} (color online) a) Upper critical fields in Mo$_3$Sb$_7$ at 4.5 and 22 kbar as a function of temperature $T$, and b) pressure dependence of $T_{SDW}$ and $T_c$. The lines are guides for the eye.}
\end{figure}

 In order to investigate pressure effect on the upper critical field, $H_{c2}$, of Mo$_3$Sb$_7$, temperature dependent resistivity data were taken at various magnetic fields and pressures. Typical results are shown in Fig. \ref{fig:Fig_3} b. The $\mu_0 H_{c2}(0)$ values deduced at  4.5 and 22 kbar are shown in Fig. \ref{fig:Fig_4}~a. Apparently, the slope of the upper critical field $dH_{c2}/dT$ for both values of applied pressure is the same near $T_c$. Werthamer et al.\cite{Werthamer} derived an expression (abbreviated as WHH model) for the upper critical field $\mu_0H_{c2}$ in terms of orbital pair-breaking, including the effect of Pauli spin paramagnetism and spin-orbit scattering. The WHH model is based on two parameters, namely, $\alpha$ the Pauli paramagnetic limitation or Maki parameter, and $\lambda_{so}$  which describes spin-orbit scattering. While the value of  $\alpha$ allows a rough discrimination between orbital pair breaking and Pauli limiting, $\lambda_{so}$ is dominant by the atomic numbers of the elements of the material under consideration. The Maki parameter can be estimated,\cite{maki} from the Sommerfeld $\gamma$ value and the residual resistivity in the normal state $\rho_0$ via $\alpha = (3e^2\hbar \gamma \rho_0)/(2m\pi^2k_B^2)$, in which \emph{e} denotes the charge and \emph{m} the mass of an electron. Considering the experimental values for $\gamma$ = 0.345 mJ/molK$^2$ and $\rho_0$ = 95 $\mu \Omega$cm results in a value of  $\alpha$= 0.71. A value for  of similar magnitude can be derived from,\cite{maki}  $\alpha$ = -0.528$\mu_0dH_{c2}/dT$ = 0.66-0.73.\cite{Tran_AM,Tran08JOP}
Setting $\alpha$ =0.71, the orbital critical field in the weak coupling limit is evaluated by $\mu_0H_{c2}(0) = 0.693T_c(-\frac{d\mu_0H_{c2}(T)}{dT})\mid_{T_c}$.\cite{Werthamer}  Thus, the difference of $\mu_0H_{2}(0)$ observed in Fig. \ref{fig:Fig_4}~a for different pressures is a result of the pressure dependent $T_c$'s.
\par The value  observed for $\alpha \sim$ 0.7 definitely indicates that orbital
pair-breaking is the essential mechanism, which limits the upper critical field.  Taking the Maki parameter $\alpha$ = 0.71 and the spin-orbit coupling $\lambda_{so}$ = 15, the overall temperature dependence of the upper critical field $\mu_0H_{c2}$ as derived from the WHH model is displayed as a dashed-dotted line in Fig. \ref{fig:Fig_4} a. As can be seen, the theoretical curves explain in a satisfying manner the experimental data revealing $\mu_0H_{c2}$ = 1.87 T for \emph{p} = 4.5 kbar and  2.05 T for \emph{p} = 22 kbar. The coherence length $\xi_0$ is calculated from $\xi_0 =\sqrt{\frac{\Phi_0}{2\pi H_{c2}(0)}}$, where $\Phi_0$ is the magnetic flux quantum. Using the theoretical $H_{c2}(0)$ values, the coherence length is estimated as 12.7 nm for 22 kbar, which is slightly smaller than the value of 13.3 nm derived for 4.5 kbar.

\par The pressure-temperature phase diagram for Mo$_3$Sb$_7$ is displayed in Fig. \ref{fig:Fig_4} b. The present data evidence that superconductivity in Mo$_3$Sb$_7$ has several remarkable features. First, superconductivity  exists in this material over a large range of applied pressures becoming essentially enhanced if pressure rises. For conventional BCS-superconductors, pressure usually suppresses superconductivity. Thus, in view of its response to the hydrostatic pressure, superconductivity in Mo$_3$Sb$_7$ is likely to be unconventional, at least different from that of metallic Al, Pb or Nb$_3$Sn. Second, there is a pressure-induced SDW transition; this SDW phase interplays with superconductivity in Mo$_3$Sb$_7$. The observed narrowing of the superconducting transition under pressure strongly emphasizes that there is a homogeneous coexistence of the SDW and the superconducting state, i.e., the same electrons are responsible for the SDW and superconductivity. This is a convincing experimental evidence for the close relationship between magnetism and the formation of Cooper pairs in the compound studied. Third, comparing the characteristic evolution of $T_{SDW}$ and $T_c$ with other SDW superconductors,\cite{SDW_SC} or with magnetic heavy-fermion superconductors,\cite{Jacard,Kastner,Tpark} various similarities are established, namely with increasing pressure, $T_{SDW}$/$T_N$ decrease, whereas $T_c$ increases. The decrease of the SDW/AF ordering temperature could imply an increase of the density of states at the Fermi energy, thus enhancing superconductivity. A direct conclusion emerges that the SDW/AF and superconducting phases are two competing phases.
\par In summary, electrical resistivity and magnetization measurements were carried out under hydrostatic pressure for Mo$_3$Sb$_7$. The application of pressure increases the superconducting phase transition temperature $T_c$. Also, pressure induces a transition from a metallic to a SDW phase above 4.5 kbar. From the pressure response of $T_{SDW}$ and $T_c$, we suggest that superconducting Mo$_3$Sb$_7$ is a unique example, where superconductivity competes with a SDW phase, which is, unexpectedly, induced by pressure. The increasing $T_c$ in presence of antiferromagnetic interactions implies that superconductivity in Mo$_3$Sb$_7$ is most likely mediated by magnetic fluctuations. Studies of either NMR or neutron scattering under pressure should bring new insights to the spin-density-wave phase scenario or possibly "hidden order" in this binary compound.

\par  V.H.T. and P.W. would like to thank the Ministry of Science and Higher Education in Poland for financial support under grant N202 082/0449.  R.T.K. and E.B. acknowledge support by the Austrian FWF, P18054 and COST P16.

\end{document}